\documentclass[
    ,final            
  ]
  {aipproc}

\layoutstyle{8x11single}
\usepackage{amsmath}
\newlength{\colw}
\newcommand{\braket}[1]{\langle#1\rangle}

\newcommand{\del}[2]{\frac{\partial#1}{\partial#2}}

\newcommand{\om}{\omega}

\newcommand{\psibar}{\overline{\psi}}

\begin{document}

\title{Quark--gluon plasma phenomenology from anisotropic lattice QCD}


\author{Jon-Ivar Skullerud}{
  address={Department of Mathematical Physics, Maynooth University,
    Maynooth, Co Kildare, Ireland}
}

\author{Gert Aarts}{
  address={Department of Physics, Swansea University, Swansea SA2 8PP,
  Wales}
}

\author{Chris Allton}{
  address={Department of Physics, Swansea University, Swansea SA2 8PP,
  Wales}
}

\author{Alessandro Amato}{
  address={Department of Physics, Swansea University, Swansea SA2 8PP,
  Wales}
}

\author{\\Yannis Burnier}{
  address={Institut de Th\'eorie des Ph\'enom\`enes Physiques, Ecole
    Polytechnique F\'ed\'erale de Lausanne, CH--1015 Lausanne, Switzerland}
}

\author{P. Wynne M. Evans}{
  address={Department of Physics, Swansea University, Swansea SA2 8PP,
  Wales}
}

\author{Pietro Giudice}{
  address={Institut f\"ur Theoretische Physik, Universit\"at
    M\"unster, D--48149 M\"unster, Germany}
}

\author{Simon Hands}{
  address={Department of Physics, Swansea University, Swansea SA2 8PP,
  Wales}
}

\author{\\Tim Harris}{
  address={School of Mathematics, Trinity College, Dublin 2, Ireland}
}

\author{Aoife Kelly}{
  address={Department of Mathematical Physics, Maynooth University,
    Maynooth, Co Kildare, Ireland}
}

\author{Seyong Kim}{
  address={Department of Physics, Sejong University, Seoul 143-747, Korea}
}

\author{Maria Paola Lombardo}{
  address={INFN--Laboratori Nazionali di Frascati, I--00044 Frascati
    (RM), Italy}
}

\author{\\Mehmet B. Oktay}{
  address={Department of Physics and Astronomy, University of Iowa,
    Iowa City, Iowa 52242, USA}
}

\author{Alexander Rothkopf}{
  address={Institut f\"ur Theoretische Physik, Universit\"at
    Heidelberg, Philosophenweg 16, D--69120 Heidelberg, Germany}
}

\author{Sin\'ead M. Ryan}{
  address={School of Mathematics, Trinity College, Dublin 2, Ireland}
}

\begin{abstract}
The FASTSUM collaboration has been carrying out simulations of
$N_f=2+1$ QCD at nonzero temperature in the fixed-scale
approach using anisotropic lattices. Here we present the status of
these studies, including recent results for electrical conductivity
and charge diffusion, and heavy quarkonium (charm and beauty)
physics.
\end{abstract}

\maketitle


\section{Introduction}

The heavy-ion collision experiments at RHIC and LHC have given a
wealth of information about QCD matter at high temperatures, and have
provided strong evidence for the existence of a deconfined state of
matter, the quark--gluon plasma (QGP).  Furthermore, it has been shown
that at the energy densities reached in these experiments, this state
of matter can be described well by nearly ideal hydrodynamics, and
must hence be strongly coupled.  A large amount of information has
also been obtained about various probes of the QGP, including
electromagnetic probes, high-energy jets and heavy quarks.

One of the outstanding challenges is to obtain a clear theoretical
understanding of all the phenomena observed in the heavy-ion
collisions.  In particular, first-principles calculations of transport
coefficients such as viscosity and conductivity would place the
hydrodynamical description on a solid footing.  Heavy quarkonia have
been proposed as ``thermometers'' of the QGP, but this requires a
precise knowledge of their dissociation rates as a function of
temperature.

Lattice QCD is the method of choice for obtaining quantitative
theoretical predictions for QCD matter at high temperature and zero or
small net baryon density.  However, to obtain information about
real-time quantities such as transport coefficients and dissociation
rates, analytical continuation of the imaginary-time information
obtained from lattice simulation is necessary.  This is in principle
an ill-posed problem, but it can be addressed using Bayesian methods
such as the maximum entropy method (MEM) \cite{Asakawa:2000tr} and the
new bayesian reconstruction method of \cite{Burnier:2013nla}, or with
alternative, model independent
methods \cite{Cuniberti:2001hm,Burnier:2012ts}.

All these methods require a fine resolution in the
temporal direction to obtain reliable results.  This may be done by
using anisotropic lattices with a smaller spacing in time than in
space.  The FASTSUM collaboration has for a number of years been
carrying out simulations on anisotropic lattices with dynamical
fermions with this in mind.  Here we present some of the most recent
results from these simulations.

\section{Simulation and analysis details}

We have generated two sets of anisotropic lattice ensembles using
improved Wilson-type fermions, see Table~\ref{tab:params} for
details.  The 1st generation ensemble
\cite{Morrin:2006tf,Oktay:2010tf} had $N_f=2$ active flavours with
$m_\pi\approx500$MeV, while the 2nd generation ensemble has $N_f=2+1$
active flavours with a physical strange quark and
$m_\pi\approx400$MeV.  The 2nd generation parameters are the same as
those used by the Hadron Spectrum Collaboration
\cite{Edwards:2008ja,Lin:2008pr}, who have kindly 
allowed us to use their zero-temperature configurations for the
purpose of these studies.  The 2nd generation configurations as well
as light and charm quark correlators were produced using the Chroma
software system \cite{Edwards:2004sx} with BAGEL optimisation
\cite{Boyle:2009vp}.

For each parameter set we have generated gauge configurations at a
range of temperatures, see table~\ref{tab:temps} for details.  Note
that the second generation includes temperatures both below and above
the deconfinement transition.  The pseudocritical temperature $T_c$ was
determined from the inflection point of the renormalised Polyakov
loop, see \cite{Allton:2014uia,Aarts:2014nba} for details.

\begin{table}
\begin{tabular}{c|ccccccc}\hline
Gen & $N_f$& $\xi$  & $a_s$ (fm) & $a_\tau^{-1}$ (GeV) & $m_\pi/m_\rho$ & $N_s$
& $L_s$ (fm)  \\ \hline
1 & 2   & 6.0 &  0.162 &   7.35 & 0.54 & 12 & 1.94 \\
2 & 2+1 & 3.5 &  0.123 &   5.63 & 0.45 & 24 & 2.94 \\
 & & & & & & 32 & 3.94 \\
\hline
\end{tabular}
\caption{Parameters for the 1st and 2nd generation ensembles.
  $\xi=a_s/a_\tau$ is the anisotropy; $N_s$ is the number of spatial
  sites and $L_s$ is the extent of the lattice in the spatial directions.}
\label{tab:params}
\end{table}

\begin{table}
\begin{tabular}{ccc|ccc}\hline
\multicolumn{3}{c}{Gen 1} &
\multicolumn{3}{|c}{Gen 2}\\
 $N_\tau$ & $T$ (MeV) & $T/T_c$ &  $N_\tau$ & $T$ (MeV) & $T/T_c$ \\\hline
 80 & \,92 & 0.42  & 128 & \,44 & 0.24 \\
 32 & 230  & 1.05  &  48 & 117  & 0.63 \\
 28 & 263  & 1.20  &  40 & 141  & 0.76 \\
 24 & 306  & 1.40  &  36 & 156  & 0.84 \\
 20 & 368  & 1.68  &  32 & 176  & 0.95 \\
 18 & 408  & 1.86  &  28 & 201  & 1.09 \\
 16 & 459  & 2.09  &  24 & 235  & 1.27 \\
   &      &        &  20 & 281  & 1.52 \\
& &                &  16 & 352  & 1.90 \\ \hline
\end{tabular}
\caption{Temporal lattice extents $N_\tau$ and temperatures $T$ used
  in our simulations.}
\label{tab:temps}
\end{table}

Transport properties and spectral information are encoded in spectral
functions $\rho(\om,\vec{p})$, which are related to the euclidean
correlators $G_E(\tau,\vec{p})$ that can be computed on the lattice
through the integral relation
\begin{equation}
G_E(\tau,\vec{p})
 = \int_0^\infty\rho(\om,\vec{p})K(\tau,\om;T)\frac{d\om}{2\pi}\,,
\quad K(\tau,\om;T) =
 \frac{\cosh[\om(\tau-\frac{1}{2T})]}{\sinh\frac{\om}{2T}}
\label{eq:spectral}
\end{equation}
for a system in thermal equilibrium.  We have used the maximum entropy
method with Bryan's algorithm and the modified kernel
proposed in \cite{Aarts:2007wj} to determine the most likely spectral
function given our correlators.  We have also employed the novel
Bayesian method developed in \cite{Burnier:2013nla}, and will present
results using this method in the beauty sector.

\section{Conductivity, charge susceptibility and charge diffusion}
\label{sec:cond}

The electrical conductivity is related to the low-frequency limit of
the current--current spectral function,
\begin{equation}
\sigma = \frac{1}{6}\lim_{\om\to0}\frac{\rho_{em}(\om)}{\om}\,,
\end{equation}
with
\begin{equation}
G^{em}_{ii}(\tau)
 = \int d^3x\braket{j_i^{em}(\tau,\vec{x})j_i^{em}(0,\vec{0})^\dagger}
 = \int_0^\infty\rho_{em}(\om)K(\tau,\om)\frac{d\om}{2\pi}\,,
\end{equation}
where $j_\mu^{em}=\sum_f(eq_f)j_\mu^f$ is the electromagnetic current
operator and $q_f$ are the quark charges in units of the elementary
charge $e$.  We have used the exactly conserved vector current on the
lattice to compute $G^{em}_{ii}(\tau)$, see \cite{Amato:2013naa,Aarts:2014nba} for
details.   In order to compare results obtained from simulations with
different flavour content, in particular results from 2 (u, d) and 3
(u, d, s) flavours, the spectral functions and conductivity have been
divided by the factor $C_{em} = \sum_f(eq_f)^2$, i.e.\ the sum of the
squares of the quark charges.

We have also computed the susceptibilities of 
baryon number, isospin and electric charge as a function of
temperature \cite{Giudice:2013fza,Aarts:2014nba}.  These are of substantial
phenomenological interest as they are related to the event-by-event
fluctuations of these quantities in heavy-ion collisions, and may be
used to locate the transition to the QGP and the critical endpoint in
the temperature -- chemical potential plane, if it exists.

Furthermore, the conductivity is directly related to the charge diffusion
coefficient $D_Q$ through the relation $D_Q=\sigma/\chi_Q$, where
$\chi_Q$ is the electric charge susceptibility.  While the analogous
diffusion coefficient for heavy quarks has been widely studied, much
less is known about diffusion of electric charge carried by light
quarks.  Here we present the first lattice calculation of this
quantity.  All the results in this section have been obtained with
our 2nd generation ensemble.

\begin{figure}[t]
\includegraphics*[width=\colw]{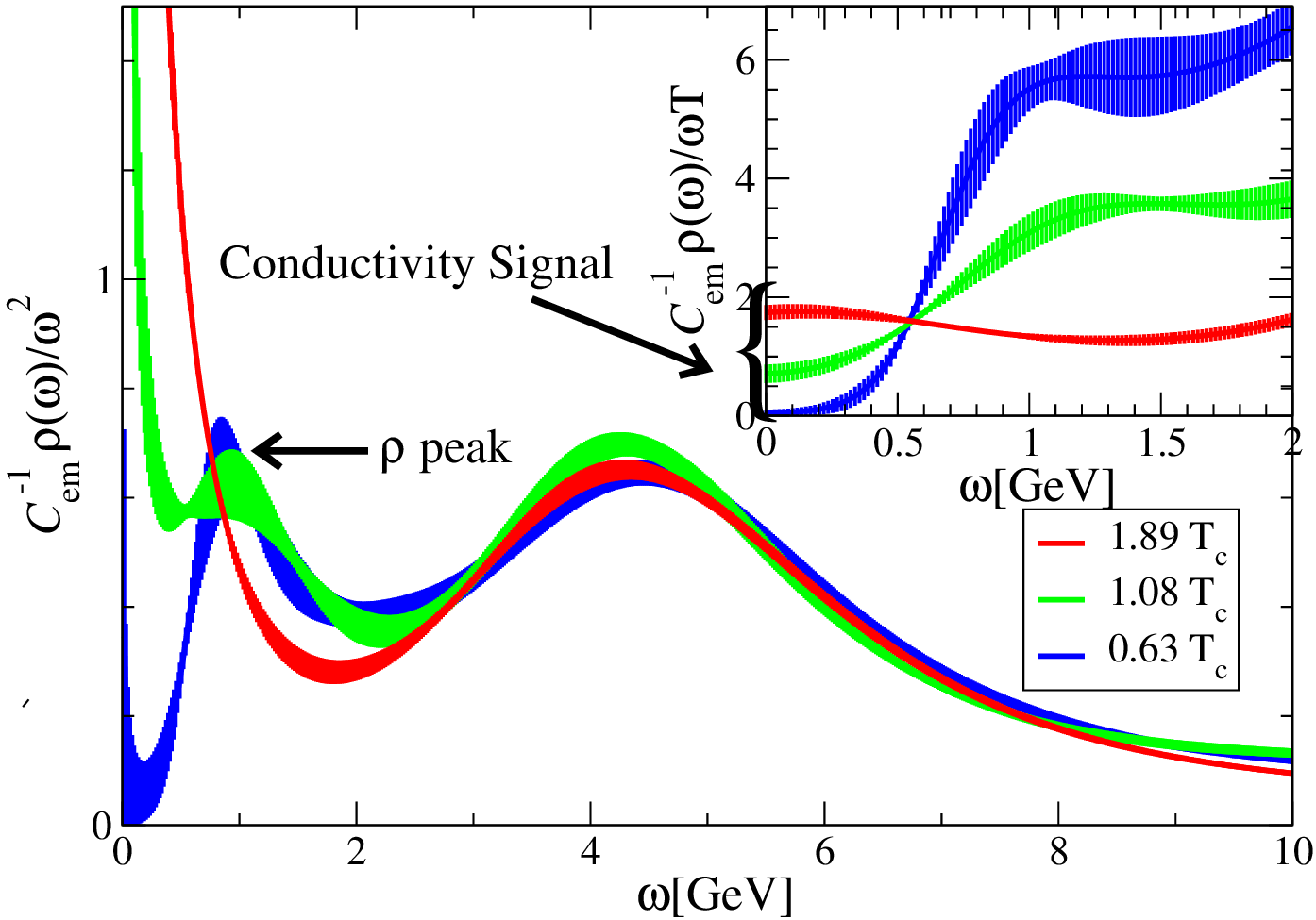}
\includegraphics*[width=\colw]{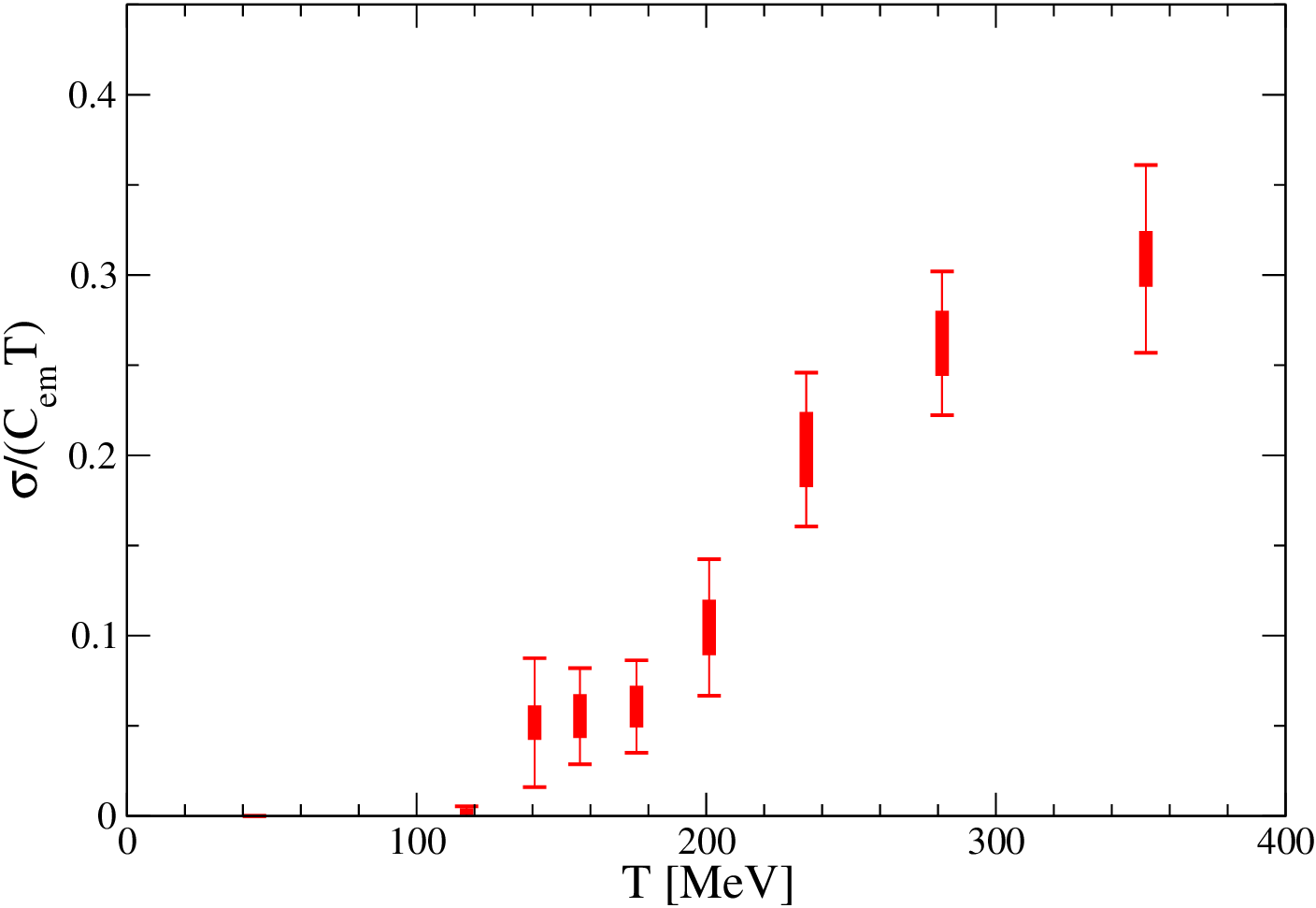}
\caption{Left: light quark vector current spectral functions for
  three different temperatures below, near and above the
  pseudocritical temperature.  The bands represent the statistical
  uncertainty.  Right: the electrical conductivity
  (including contributions from u, d and s quarks) as a function of
  temperature.  The boxes represent the uncertainties due to
  variations in the default model, while the error bars represent the
  total (systematic and statistical) uncertainty.} 
\label{fig:conductivity}
\end{figure}

In the left hand panel of figure~\ref{fig:conductivity} we show the
light (u, d) quark contribution to $\rho_{em}(\om)$ as a function of
$\om$ for three different temperatures: below, near and above the
pseudocritical temperature $T_c$.  Below $T_c$ we can clearly identify
the $\rho$ meson peak in the spectral function, while this peak
disappears at higher temperatures, in accordance with the $\rho$ meson
no longer being bound.  Above $T_c$ we find that $\rho(\om)/\om$ has a
nonzero intercept with the $\om=0$ axis, signalling a nonzero value
for the conductivity.  The MEM analysis has been carried out using a
default model $m(\om)=m_0\om(b+\om)$, which allows for a finite,
nonzero transport contribution while reproducing the continuum free
spectral function at large $\om$.

The resulting values for the conductivity are shown in the right hand
plot of figure~\ref{fig:conductivity}.  We have studied the stability of
our results with respect to variations in the default model parameter
$b$; the resulting systematic uncertainties are represented by the
filled boxes in the figure, while the error bars represent the total
(systematic and statistical) uncertainties.  We have also investigated the
stability of our results with respect to other systematics including
the range and number of time slices included, and found that they are
stable in all cases.


\begin{figure}[tb]
\includegraphics*[width=\colw]{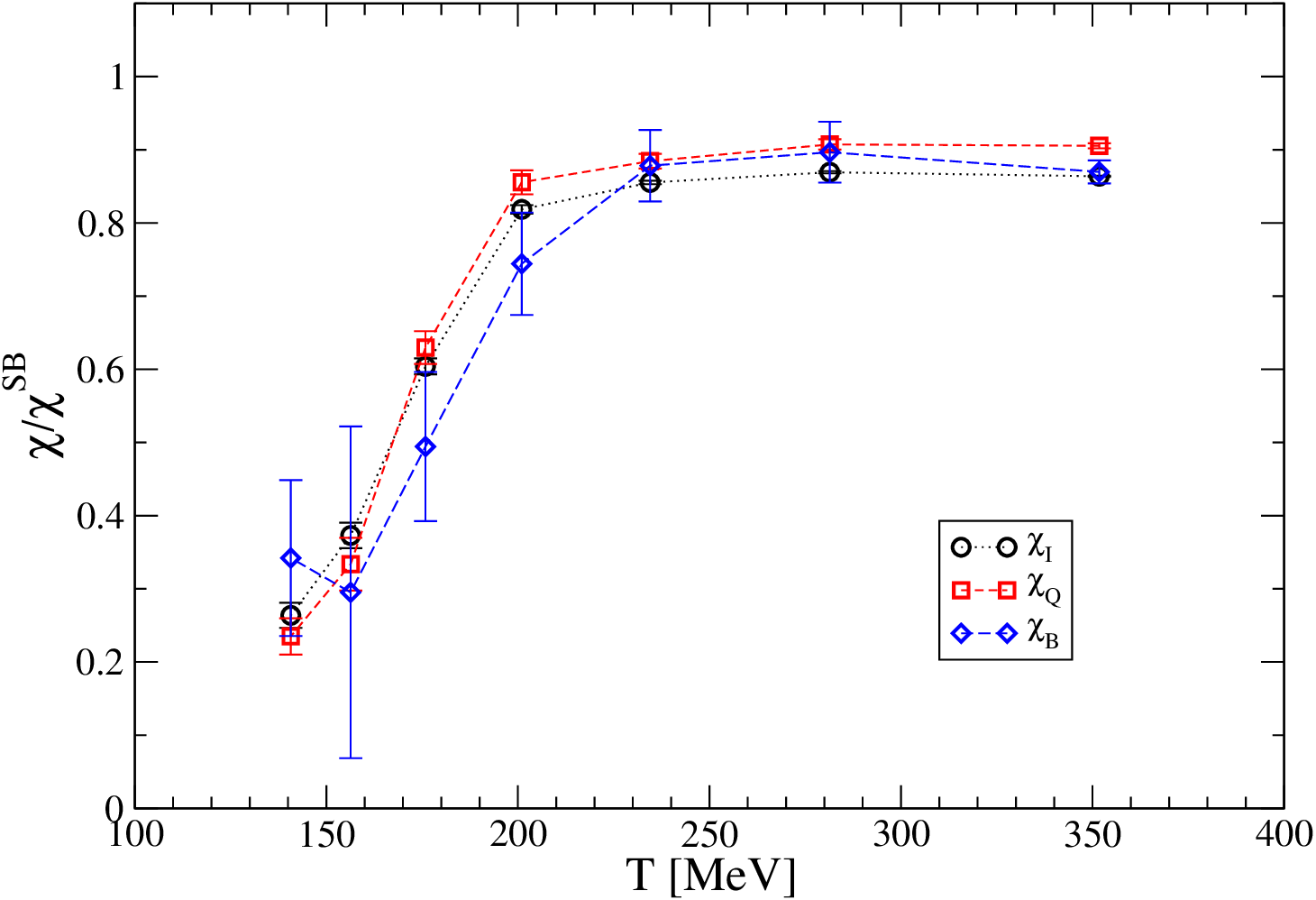}
\includegraphics*[width=\colw]{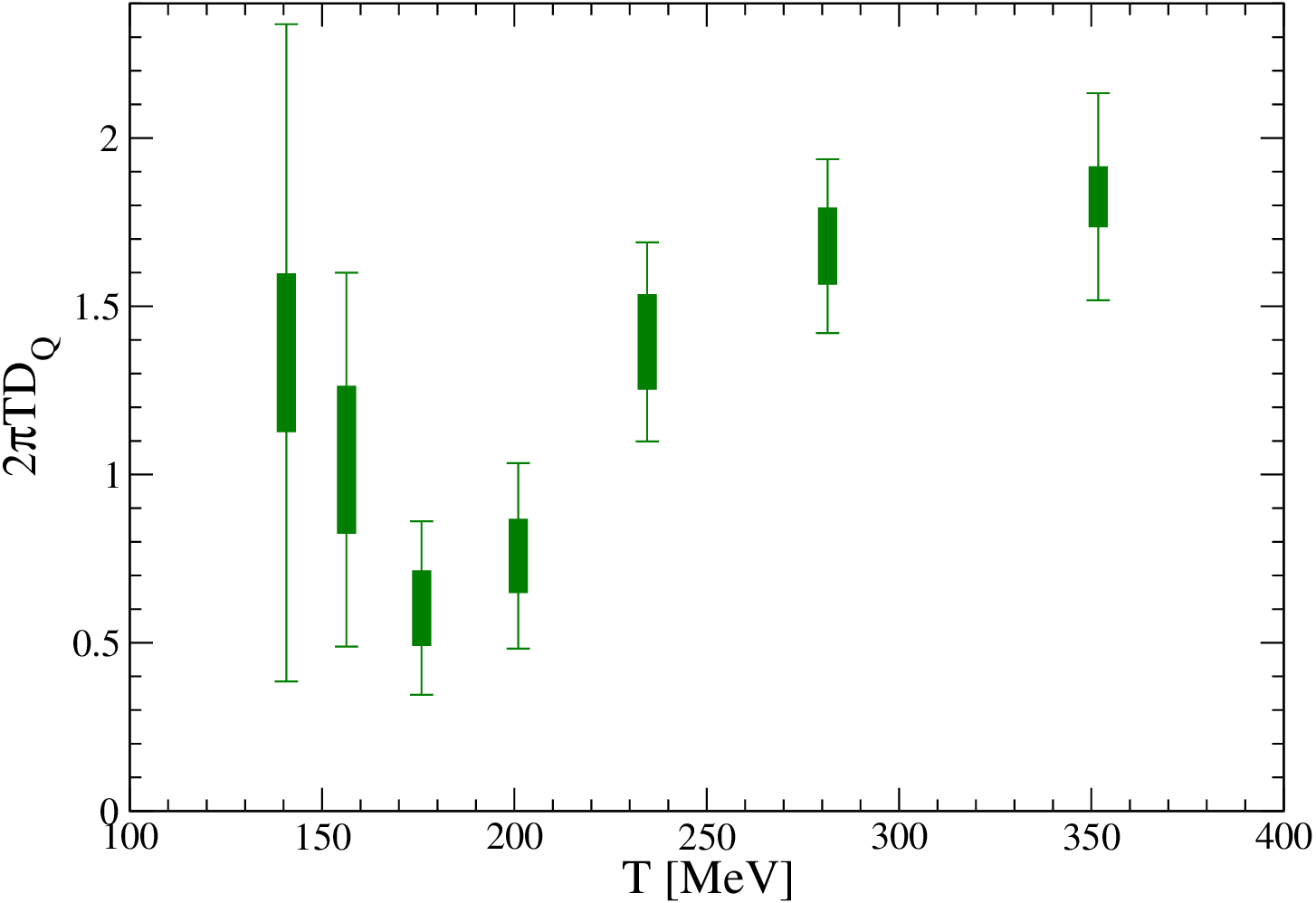}
\caption{Isospin, electric charge and baryon susceptibility $\chi_I,
  \chi_Q, \chi_B$, normalised by their
  values $\chi^{SB}$ in the noninteracting case (left) and charge
  diffusion coefficient $D_Q$ (right) as function of temperature.
  These results include the contributions from u, d and s quarks.  The
  meaning of the filled boxes and error bars are the same 
as in figure~\ref{fig:conductivity}.}
\label{fig:suscQdiffQ}
\end{figure}

In the left panel of figure~\ref{fig:suscQdiffQ} we show our results
for the isospin, electric charge and baryon susceptibilities as
functions of temperature.  They all show a similar behaviour,
increasing rapidly near the crossover temperature $T_c$ and
approaching the value for a gas of free quarks and gluons at high
temperature, as expected.

The right panel of figure~\ref{fig:suscQdiffQ} shows the charge
diffusion coefficient $D_Q$ as a function of temperature.  Our results
suggest that it has a minimum near $T_c$, with a value close to that
predicted in AdS/CFT models, $D_Q=1/(2\pi T)$.

\section{Charm}
\label{sec:charm}

Charmonium has been one of the most intensely studied probes in
heavy-ion collisions since charmonium suppression was proposed as a
signature of the QGP by Matsui and Satz \cite{Matsui:1986dk}.  There
have been a number of lattice calculations of charmonium spectral
functions, mostly in the quenched approximation
\cite{Asakawa:2003re,Umeda:2002vr,Datta:2003ww,Ding:2012sp}, but also
recently using 2+1 light flavours \cite{Borsanyi:2014vka}.  We have
previously studied charmonium on our 1st generation ($N_f=2$)
ensembles \cite{Aarts:2007pk,Oktay:2010tf}; here we will present
results from our 2nd generation ensembles, as well as updated results
for the momentum dependence of charmonium correlators from the 1st
generation ensembles.

\subsection{Charmonium at zero and nonzero momentum}

\begin{figure}[tb]
\includegraphics*[width=\colw]{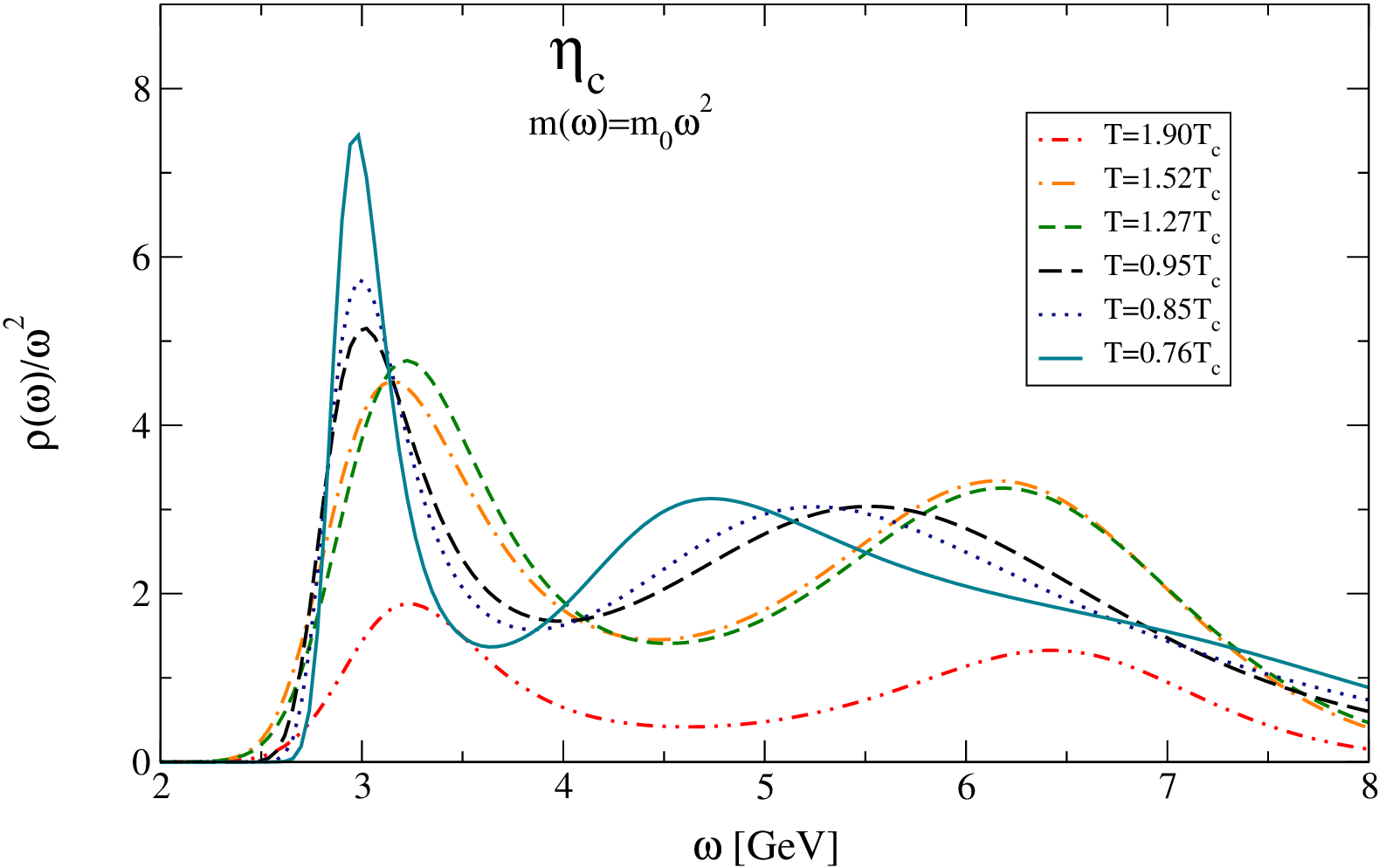}
\includegraphics*[width=\colw]{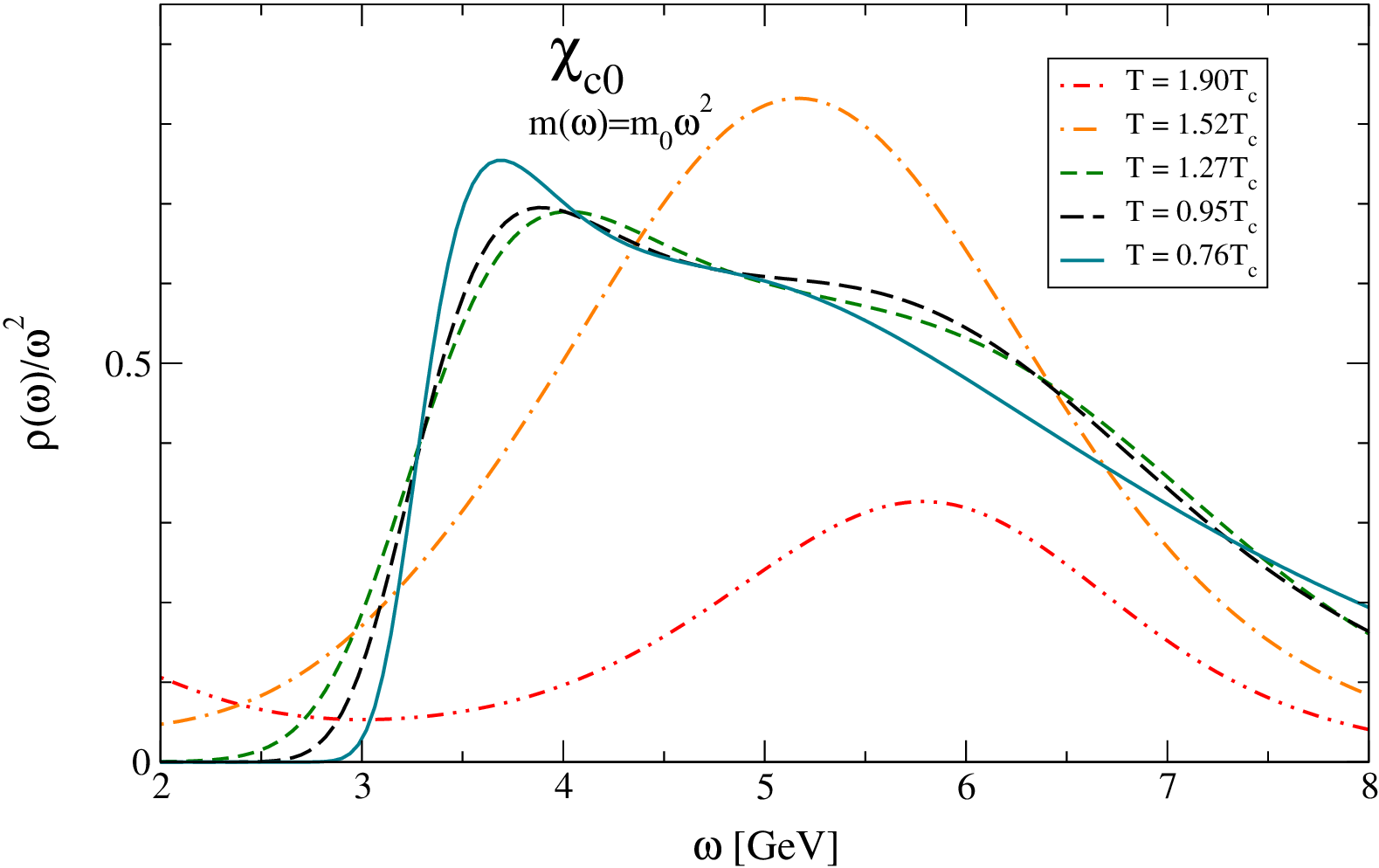}
\caption{Charmonium spectral functions from the second generation
  ensemble.  Pseudoscalar channel on the left, scalar channel on the
  right.}
\label{fig:charm-gen2}
\end{figure}

Figure~\ref{fig:charm-gen2} shows charmonium spectral functions in the
pseudoscalar (S-wave) and scalar (P-wave) channels from our second
generation ensembles.  Our results suggest that the S-wave ground
state survives up to $T\approx1.6T_c$, while the P-wave ground state
melts close to $T_c$.  This is consistent with our previous first
generation results \cite{Aarts:2007pk}.  We have studied the
dependence of these results on the default model used, and found that
the results are stable for all but the highest temperature.  We are in
the process of analysing the data using the alternative Bayesian
method of \cite{Burnier:2013nla}.

\begin{figure}[tb]
\includegraphics*[width=0.9\textwidth]{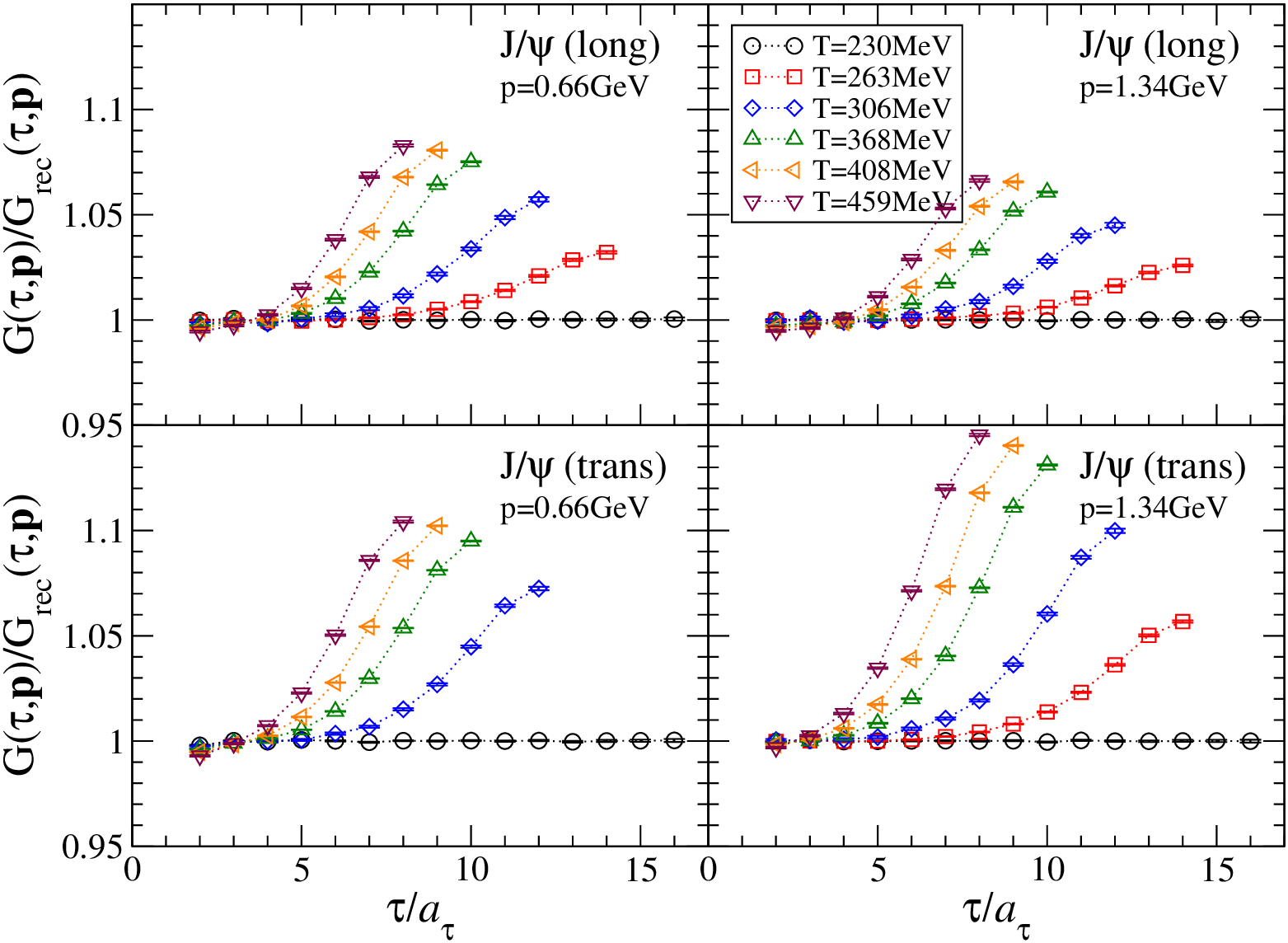}
\caption{Reconstructed charmonium correlators in the longitudinal
  (top) and transverse (bottom) vector channel (1st generation), at
  two different momenta.  The reference temperature is $T_r=230$MeV
  ($N_\tau=32)$.} 
\label{fig:charm-mom}
\end{figure}

At nonzero momentum, the MEM reconstruction has much larger systematic
uncertainties than at zero momentum \cite{Oktay:2010tf,Kelly:2013cpa}.
To circumvent this problem, we may instead compare the correlators
at temperature $T$ with the \emph{reconstructed correlators} which result from
integrating \eqref{eq:spectral} using the kernel evaluated at $T$, but
with a spectral function
$\rho(\om,\vec{p};T_r)$ determined at a reference temperature $T_r$.
The reference temperature is chosen to be relatively low so that
the MEM determination of $\rho(\om,\vec{p},T_r)$ is under control.
Figure~\ref{fig:charm-mom}
shows this comparison for the vector channel at two different momenta
with $T_r=230$MeV.  Note that the comparison is always made with the
reconstructed spectral function for the same momentum.  We see that
the combined effect of the temperature and momentum is different in
the longitudinal and transverse channel: while the thermal
modifications in the longitudinal channel become smaller with
increasing momentum, in the transverse channel they increase with the
momentum, and are also larger in magnitude.

\subsection{Charmonium potential}

Nonrelativistic potential models have been widely used to study bound
states of charm and beauty quarks both at zero and nonzero
temperature.  While it has long been known how their use can be given
a firm foundation at
zero temperature in effective field theory (pNRQCD), it is only
recently that a similar understanding has begun to be developed in the
high-temperature case \cite{Burnier:2007qm,Brambilla:2008cx}.  Most if
not all high-temperature applications 
have so far been based on the potential between infinitely heavy
(static) quarks, and it is not yet known how a finite quark mass will
modify this.

\begin{figure}[tb]
\includegraphics*[width=\colw]{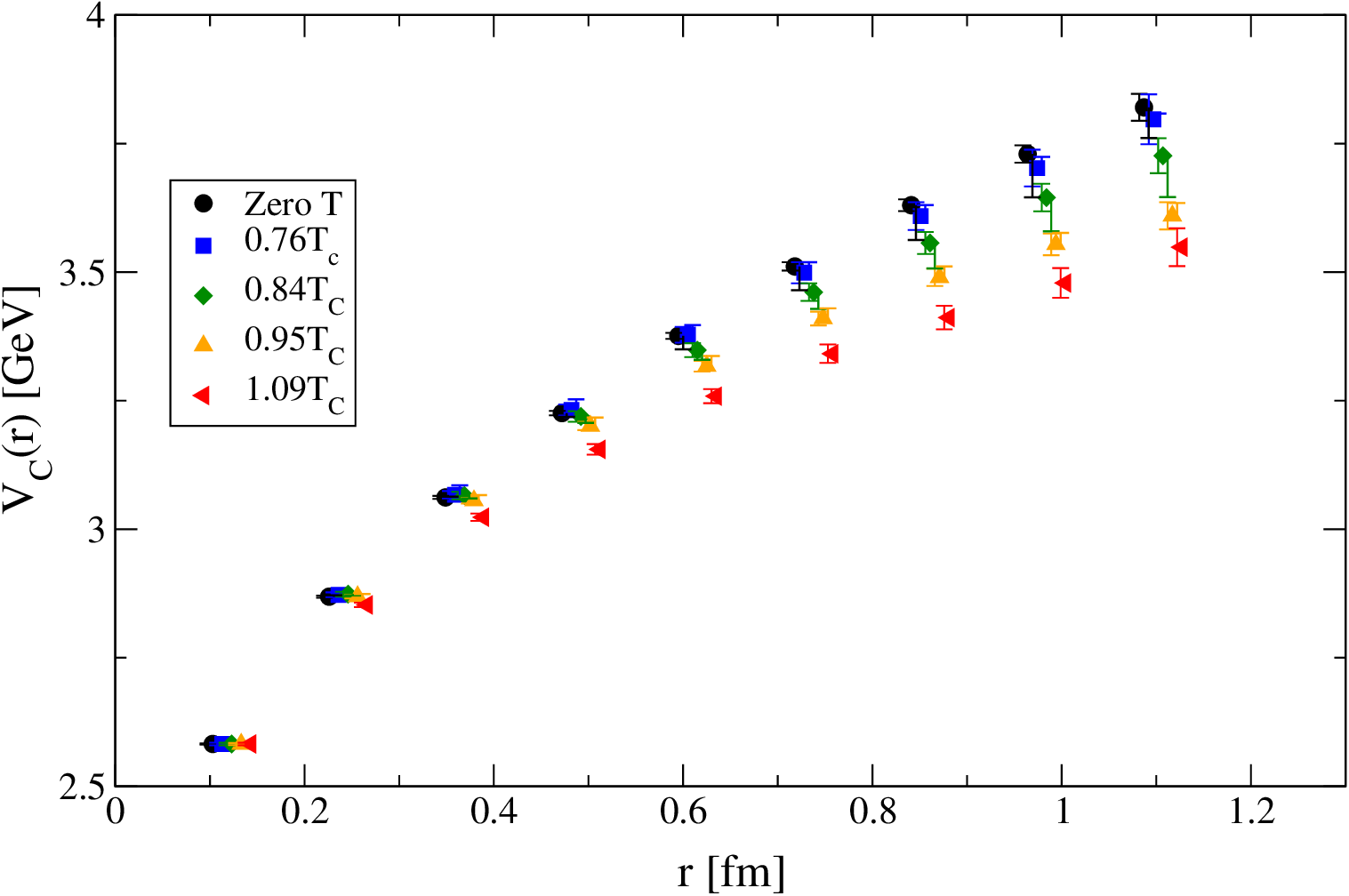}
\includegraphics*[width=\colw]{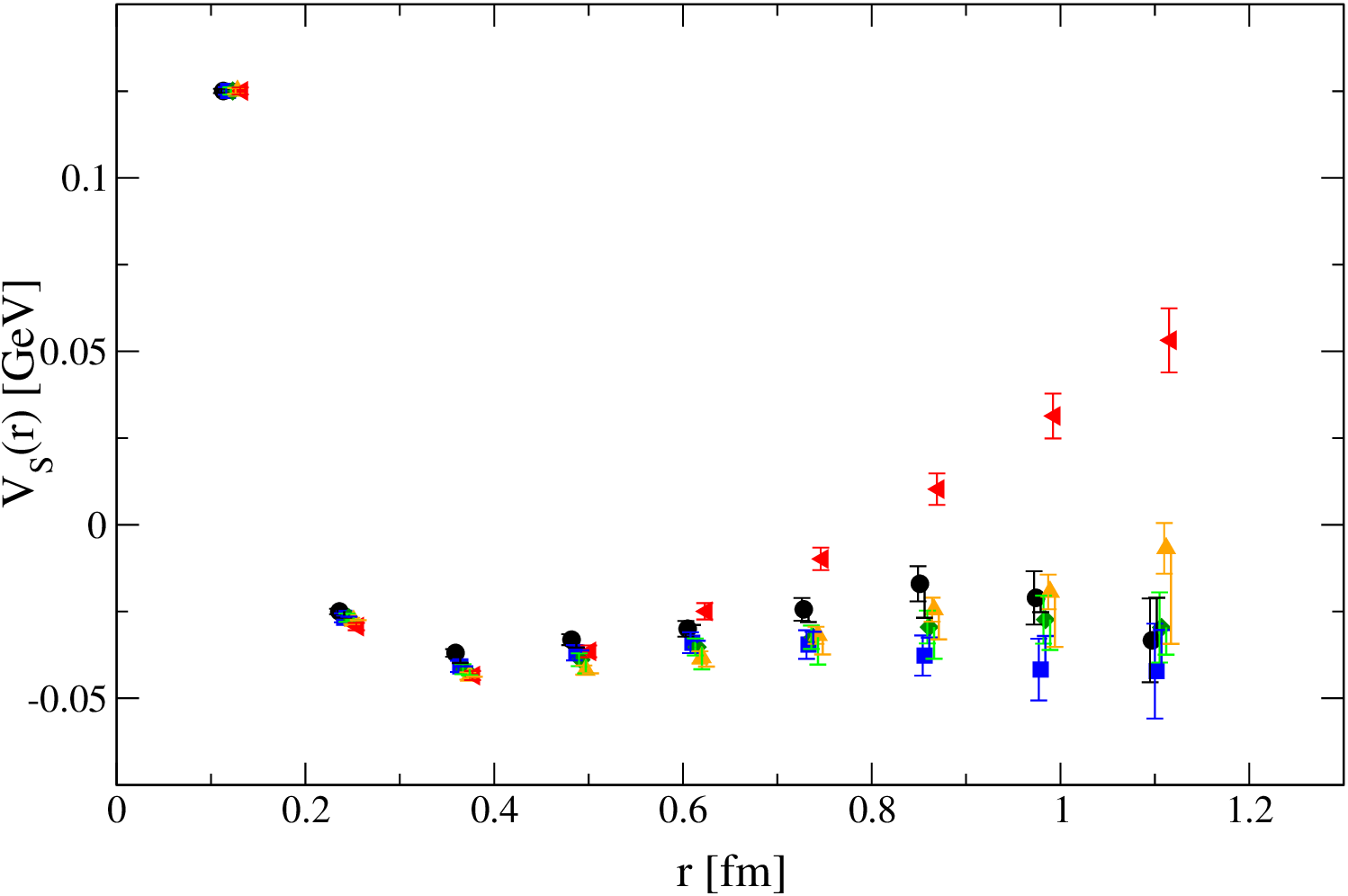}
\caption{Charmonium potential.  Left: central (spin-independent)
  potential; right: spin-dependent potential.  The points are
  horizontally shifted for clarity.  The two sets of error bars (to
  the left and right of the symbols) denote
statistical and systematic (from variations in the time range used)
uncertainties, respectively.}
\label{fig:charm-pot}
\end{figure}

We will take an alternative approach, assuming that the
charmonium system can be described using a Schr\"odinger equation with
a real potential, and determining the potential from the charmonium
correlators that we compute on the lattice.  Specifically, we compute 
point-split correlators with an operator $\Gamma=\{\gamma_\mu,\gamma_5\}$
\begin{equation}
C_\Gamma(\vec{r},\tau)
 = \sum_{\vec{x}}\braket{\psibar(\vec{x},\tau)U(x,x+\vec{r})
                  \Gamma\psi(\vec{x}+\vec{r},\tau)
                  \psibar(\vec{0},0)\Gamma^\dagger\psi(\vec{0},0)}\,.
\end{equation}
The potential $V_\Gamma(r)$ can then be determined using \cite{Aoki:2012tk}
\begin{equation}
\del{C_\Gamma(r,\tau)}{\tau}
 = \bigg(\frac{1}{m_c}\del{^2}{r^2}-V_\Gamma(r)\bigg)C_\Gamma(r,\tau)\,,
\end{equation}
which is valid for $\tau\ll1/2T$.  The vector and pseudoscalar
potentials can be combined to produce the central (spin-independent)
and spin-dependent potentials $V_C$ and $V_S$,
\begin{equation}
V_C(r) = \frac{3}{4}V_{\gamma_\mu}(r) + \frac{1}{4}V_{\gamma_5}(5)\,,\quad
V_S(r) = V_{\gamma_\mu}(r) - V_{\gamma_5}(r)\,.
\end{equation}
The results are shown in figure~\ref{fig:charm-pot}.  At low
temperatures the central potential is well represented by the
confining Cornell (linear + Coulomb) potential, while above $T_c$ it
becomes screened.  The spin-dependent potential is attractive at
intermediate distances with indications of a repulsive core.  There
appears to be a strong temperature dependence, also seen in our first
generation results \cite{Evans:2013yva}, which at present we have no
clear understanding of.

\section{Beauty}
\label{sec:beauty}

In recent years there has been an increased interest in $b$ physics at
high temperature and in heavy-ion collisions.  In part this is because
the energies reached at the LHC are such that $b\bar{b}$ pairs are created
abundantly, whereas at RHIC and SPS they were relatively rare.
Indeed, one of the early headline results from the heavy-ion
collisions at CERN was the observation of sequential $\Upsilon$
suppression by the CMS collaboration \cite{Chatrchyan:2012lxa}.

Theoretically, the beauty system provides a cleaner probe of the QGP
than the charm system, since effects such as regeneration and cold
nuclear matter effects which may obscure the interpretation of the
yields are much less significant.  In addition, while the
nonrelativistic approximation is marginal for charm quarks, it is
clearly valid for bound states of beauty quarks.  This allows us to
use non-relativistic QCD (NRQCD), which is an effective theory
obtained by integrating out the largest scale in the system, the heavy
quark mass.  In this case the kernel $K$ in \eqref{eq:spectral}
simplifies to 
\begin{equation}
K(\om,\tau) = e^{-\om\tau}\,.
\end{equation}
Note that the heavy quark is explicitly not in thermal equilibrium
here, and hence thermal (periodic) boundary conditions are not
imposed.  This has the additional advantage of doubling the number of
independent points in the temporal direction.

We have previously applied the NRQCD formalism to our first generation
ensembles in a series of papers
\cite{Aarts:2010ek,Aarts:2011sm,Aarts:2012ka,Aarts:2013kaa}.  Here we
will show results from our second generation ensembles
\cite{Aarts:2014cda}.  The details of the simulation including the
NRQCD action and the MEM reconstruction are presented in
\cite{Aarts:2014cda}.  


\begin{figure}[tb]
\includegraphics*[width=0.45\textwidth]{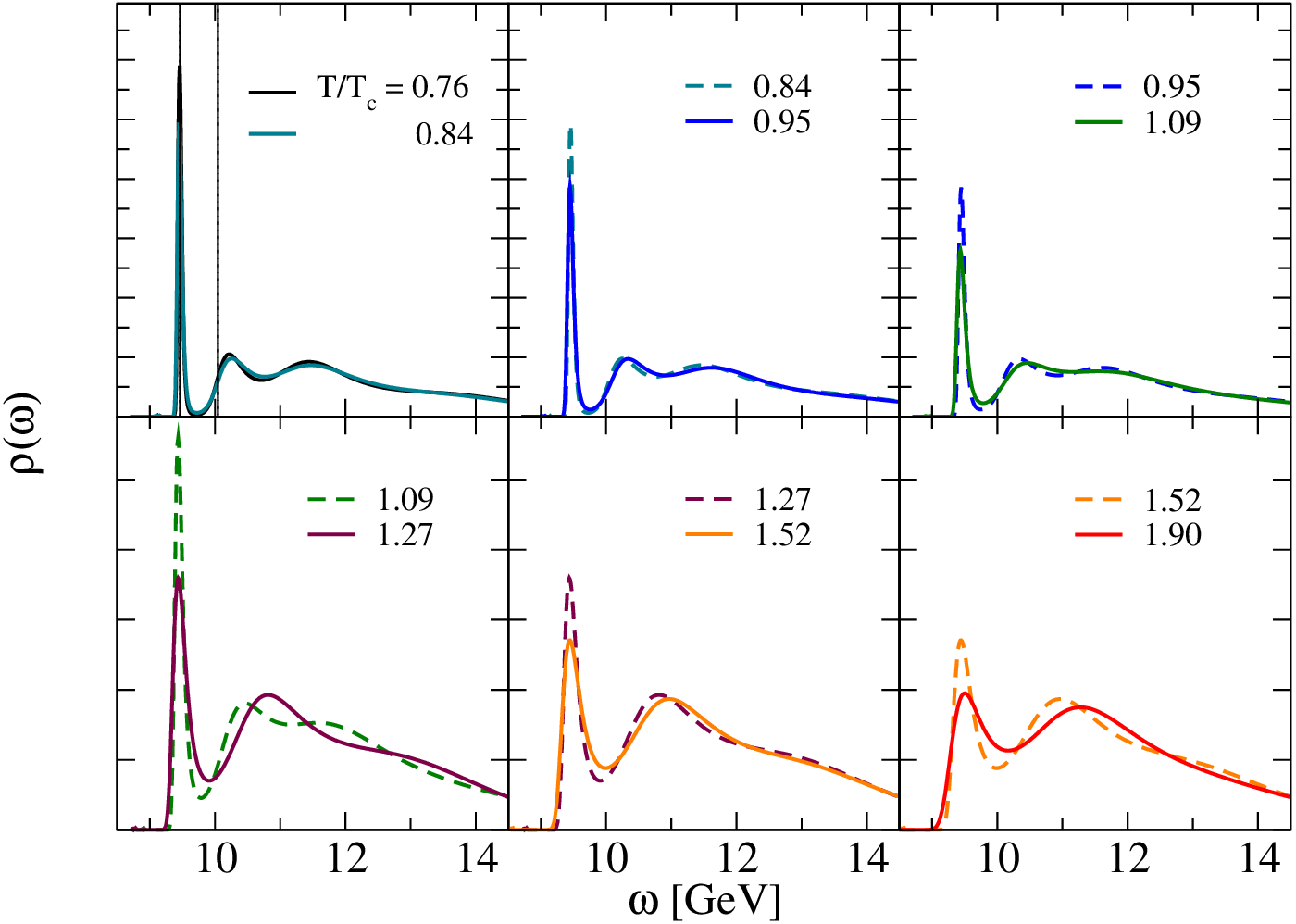}
\includegraphics*[width=0.45\textwidth]{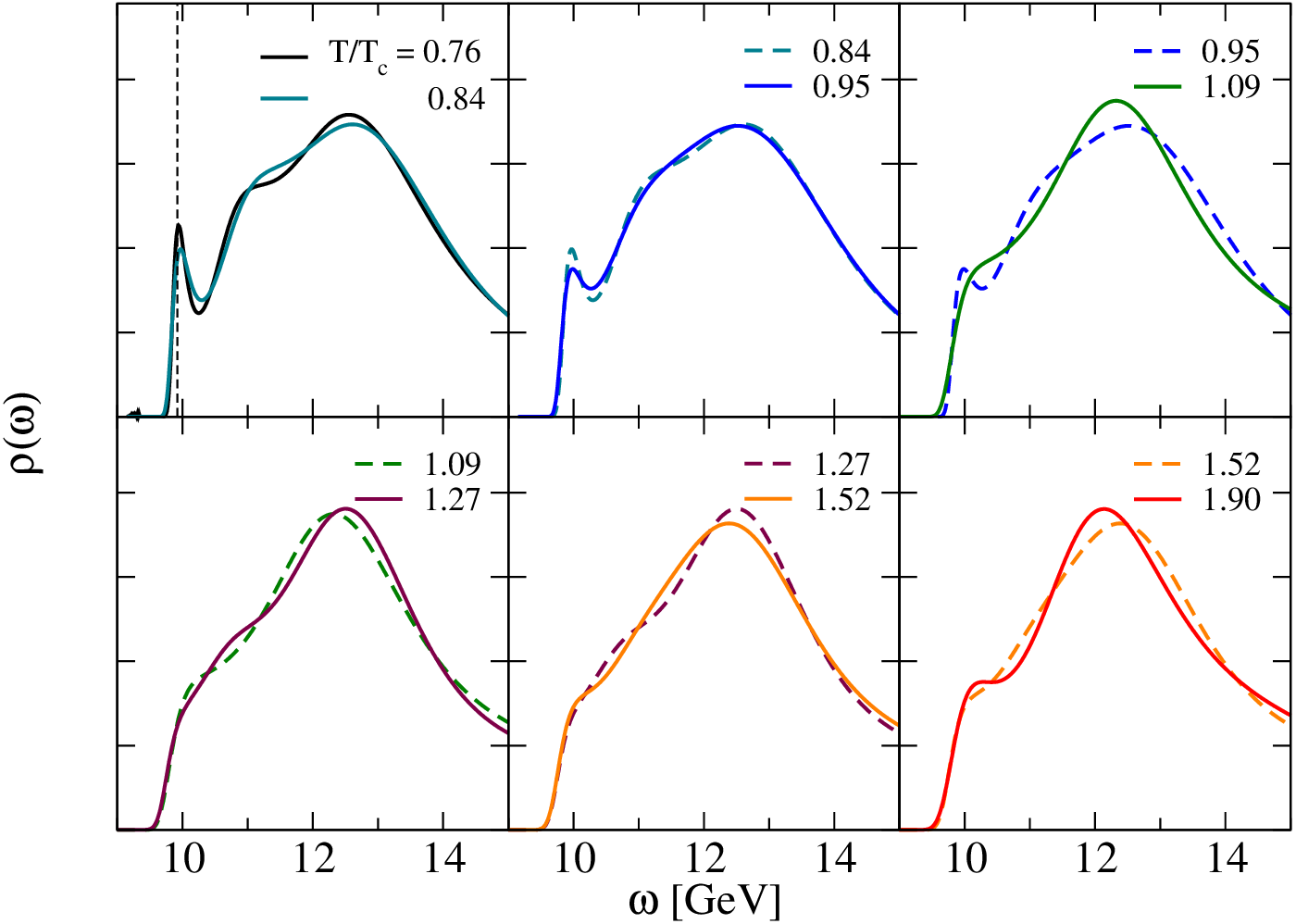}
\caption{Beautonium spectral functions in the vector channel (left)
  and axial-vector channel (right), using the maximum entropy method
\cite{Aarts:2014cda}.}
\label{fig:beauty-mem}
\end{figure}

\begin{figure}[tb]
\includegraphics*[width=0.45\textwidth]{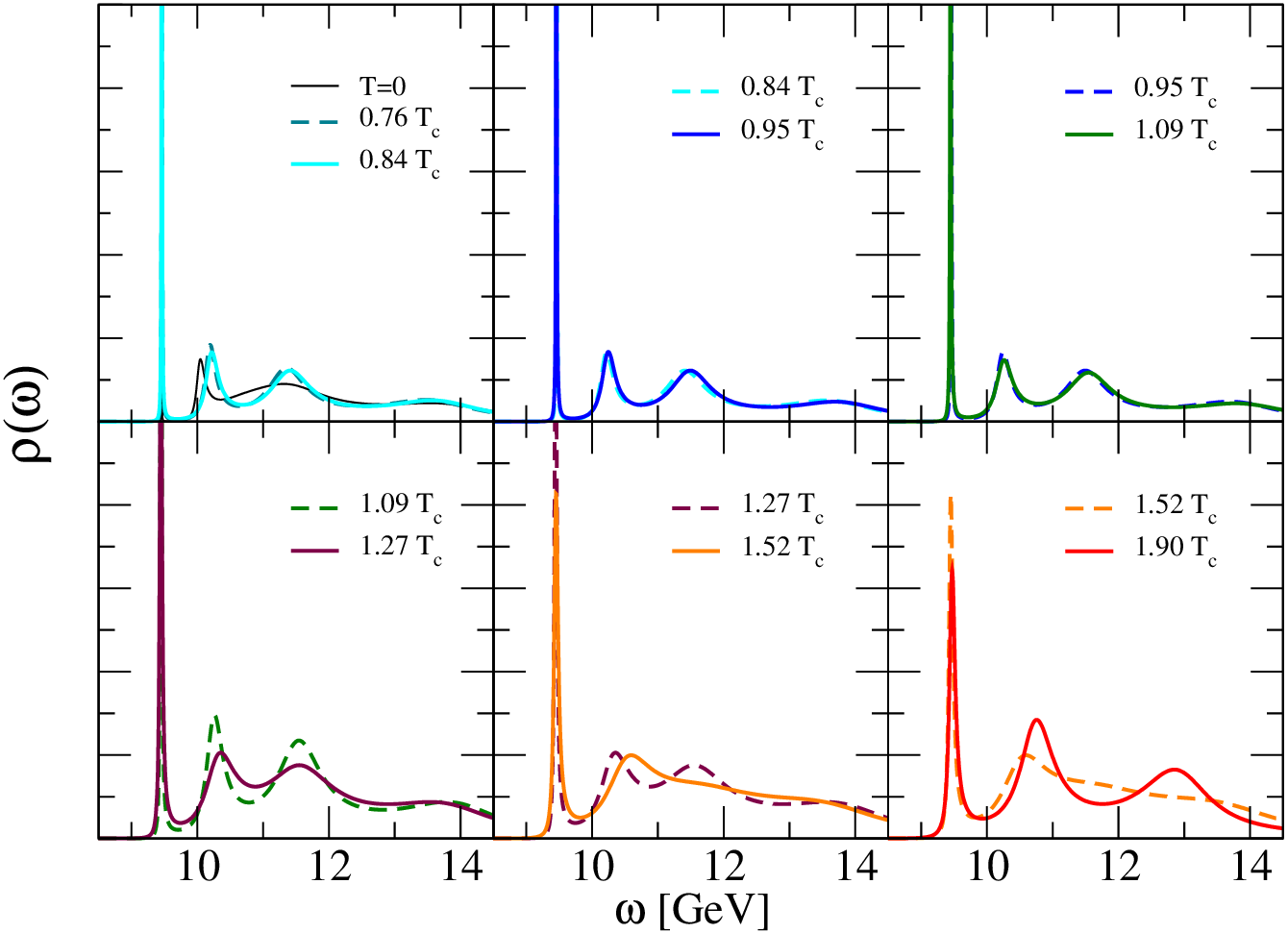}
\includegraphics*[width=0.45\textwidth]{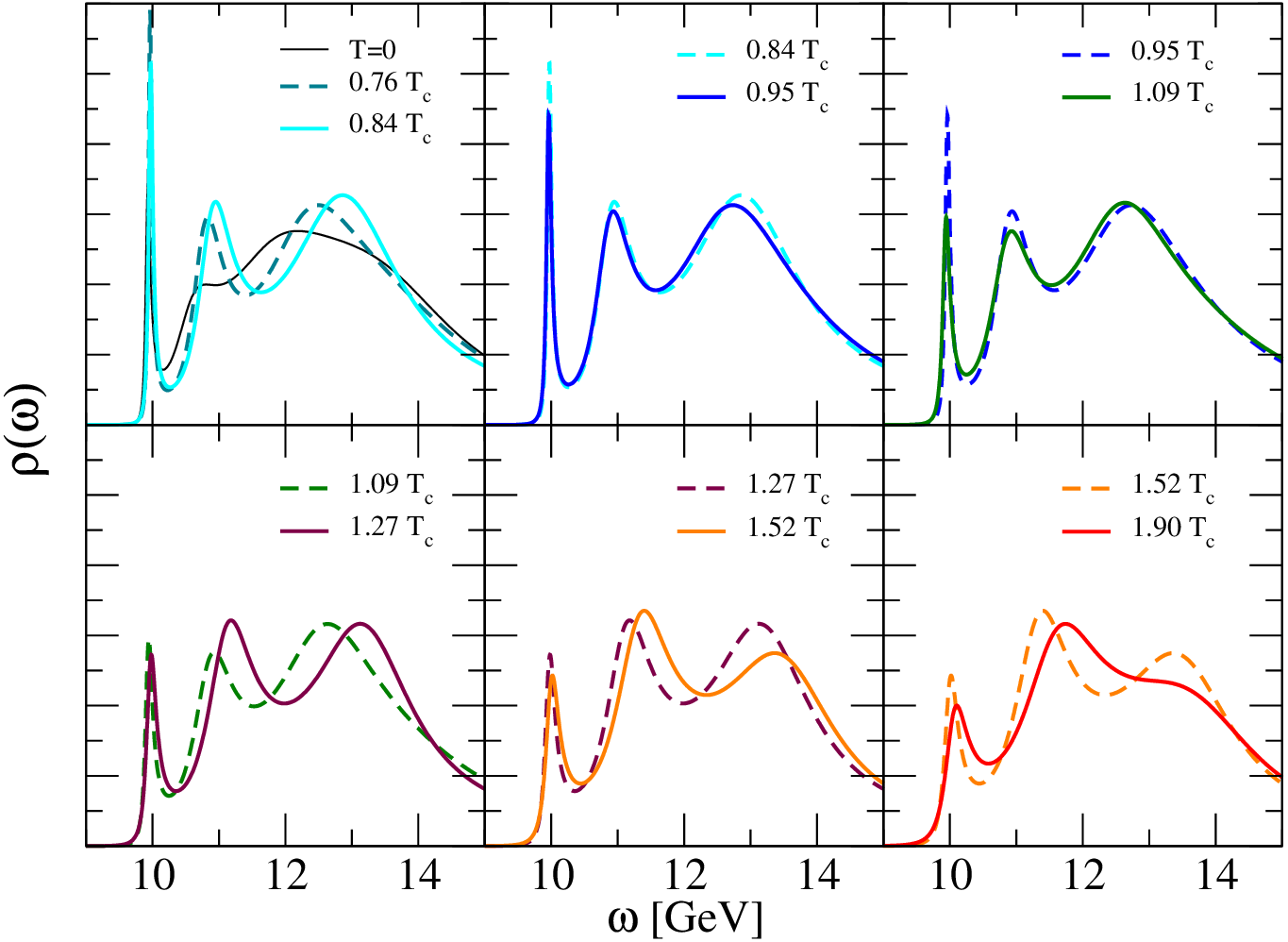}
\caption{Beautonium spectral functions in the vector channel (left)
  and axial-vector channel (right), using the novel Bayesian method of
\cite{Burnier:2013nla}.}
\label{fig:beauty-BR}
\end{figure}

Our results \cite{Aarts:2014cda} using the standard Bryan's
implementation of MEM are shown in figure~\ref{fig:beauty-mem}. We see
that the S-wave (vector) ground state $\Upsilon$ survives up to the highest
temperatures studied, while there is no evidence for any surviving
P-wave (axial-vector) ground state for temperatures above $T_c$.  This is consistent
with our results from the first generation ensembles
\cite{Aarts:2011sm,Aarts:2013kaa}.

Figure~\ref{fig:beauty-BR} shows our preliminary results using the
novel Bayesian method of \cite{Burnier:2013nla}.  For the vector
(S-wave) channel the results are in qualitative agreement with those
from the maximum entropy method, although there are quantitative
differences in the width of the ground state peak.  For the
axial-vector (P-wave) channel, on the other hand, the results in
figure~\ref{fig:beauty-BR} suggest that the ground state survives until
well into the QGP phase.  This is in accordance with the results
obtained in an analogous study using HotQCD ensembles
\cite{Kim:2014iga}.  We are currently investigating the source and
significance of these differences, and this is discussed further in
\cite{Harris:lat14}.



\section{Summary and outlook}

We have presented results from the FASTSUM collaboration's studies of
high-temperature QCD using anisotropic lattices.  Results have been
obtained in the light quark, charm and beauty sectors, including the
first results for the electrical conductivity and charge diffusion as
a function of temperature below and above the phase transition, as
well as charmonium and beautonium S and P wave states and the
potential between charm quarks.

In the beauty sector, we are currently working on a detailed
comparison of the two different Bayesian methods used in the spectral
reconstructions in order to fully understand the differences between
them.  We are also looking at alternative methods which can complement
this understanding.  We hope that this will lead to quantitative
results for the temperature-dependent mass shift and width of
$\Upsilon$ states, which will assist in the interpretation of results
from heavy-ion collisions.
In the charm sector, we have obtained preliminary results for D mesons
at high temperature, and we are also planning to study charm diffusion
using the same methods as for the conductivity and charge diffusion
(see \cite{Kelly:2013cpa} for preliminary results).  
We are also investigating baryons at high temperature, as well as the
real-time static quark potential using the methods of
\cite{Burnier:2013nla,Burnier:2014ssa}.

Finally, we are in the process of generating new ensembles with the
same quark masses and spatial lattice spacing, but with a smaller
temporal lattice spacing.  The finer temporal resolution will lead to
a more reliable spectral reconstruction and bring some of the main
systematic uncertainties associated with this under control.  It will
also allow us to reach higher temperatures, and will be the first step
towards the aim of providing quantitative predictions for spectral and
transport properties of the QGP, that is results in the continuum
limit with physical or near-physical quark masses.


\begin{theacknowledgments}
 This work is undertaken as part of the UKQCD collaboration and the
 STFC funded DiRAC Facility.  We acknowledge the PRACE grants
 2011040469 and Pra05\_1129, European Union Grant Agreement number
 238353 (ITN STRONGnet), the Irish Centre for High-End Computing,
 SFI grants 08-RFP-PHY1462 (JIS) and 11-RFP.1-PHY-3201 (TH and SMR),
 IRC (AK), Swansea University (AA), STFC, the Wolfson Foundation (GA), the
 Royal Society (GA), the Leverhulme Trust (GA and CA), SNF grant
 PZ00P2--142523 (YB) and the Research Foundation of Korea, grant
 No.\ 2010-002219 (SK) for support.
\end{theacknowledgments}



\bibliographystyle{aipproc} 

\bibliography{lattice,hot,jis}

\IfFileExists{\jobname.bbl}{}
 {\typeout{}
  \typeout{******************************************}
  \typeout{** Please run "bibtex \jobname" to optain}
  \typeout{** the bibliography and then re-run LaTeX}
  \typeout{** twice to fix the references!}
  \typeout{******************************************}
  \typeout{}
 }

\end{document}